\documentclass[aps,prb,twocolumn,floatfix]{revtex4-1}

\usepackage{graphicx,amsfonts}
\usepackage{bm,color}

\usepackage{amssymb,amsmath,hyperref}

\begin{document}

\title{Thermally pumped  on-chip maser}
\author{George Thomas$^1$}
%\affiliation{QTF Centre of Excellence, Department of Applied Physics, Aalto University, P.O. Box 15100, FI-00076 Aalto, Finland}
\author{Azat Gubaydullin$^1$}
%\affiliation{QTF Centre of Excellence, Department of Applied Physics, Aalto University, P.O. Box 15100, FI-00076 Aalto, Finland}
\author{Dmitry S. Golubev$^1$}
%\affiliation{QTF Centre of Excellence, Department of Applied Physics, Aalto University, P.O. Box 15100, FI-00076 Aalto, Finland}
\author{Jukka P. Pekola$^{1,2}$}
\affiliation{$^1$Pico group, QTF Centre of Excellence, Department of Applied Physics, Aalto University, P.O. Box 15100, FI-00076 Aalto, Finland} 
\affiliation{$^2$Moscow Institute of Physics and Technology, 141700 Dolgoprudny, Russia}

\begin{abstract}
We present a theoretical model of an on-chip three-level maser in a superconducting circuit based on a single artificial atom 
and pumped by a temperature gradient between thermal baths coupled to different interlevel transitions. 
We show that maser powers of the order of a few femtowatts, well exceeding the resolution of the sensitive bolometry,
can be achieved with typical circuit parameters. 
We also demonstrate that population inversion in the artificial atom can be detected without measuring 
coherent radiation output of the maser. For that purpose, the system should operate as a three-terminal heat transport device. 
The hallmark of population inversion is the influx of heat power into the weakly coupled output terminal even though 
its temperature exceeds the temperatures of the two other terminals.
The proposed method of on-chip conversion of heat into microwave radiation and  control of energy-level populations  
by heating  provide additional useful tools for circuit quantum electrodynamics experiments.
\end{abstract}

\maketitle

\section{Introduction}

In recent years, considerable progress has been achieved in  both 
circuit quantum electrodynamics (cQED)\cite{Wallraff,Devoretap} and circuit quantum thermodynamics \cite{pekola2015, Hoffer2016,vinjanampathy_quantum_2016}.
The experiments with superconducting quantum systems have led to spectacular fundamental studies of dynamical Casimir effect \cite{Wilson}, 
Lamb shift \cite{Fragner,Silveri}, etc.
Moreover, in cQED devices lasing and other quantum optical phenomena \cite{Haroche} can be more easily exploited 
in the strong coupling regime \cite{Wallraff, Devoretap}. 
For example, single atom lasing, which requires strong coupling, and  
which has been  demonstrated in experiments with a cavity containing 
a single rubidium \cite{Meschede1985} or cesium \cite{Kimble2003}  atom, 
has also been achieved in cQED experiments with a microwave resonator coupled to a superconducting artificial atom \cite{Astafiev2007,Astafiev2015}.
Lasing has also been observed in an alternative cQED setup with a voltage biased Josephson junction and a resonator \cite{Cassidy}.
Microwave masers based on superconducting elements and pumped by microwave sources have been theoretically analyzed, e.g., 
in Refs. \cite{Hekking2010,Nori2007,Nori2009}, and those based on semiconducting quantum dots --- in Refs. \cite{Lukin2004,Marthaler2011,Rastelli2019,Segal2019}.
Furthermore, Josephson cascade micromaser has also been proposed \cite{Kurihara1996}.

\begin{figure}
\begin{center}
\includegraphics [width=8cm] {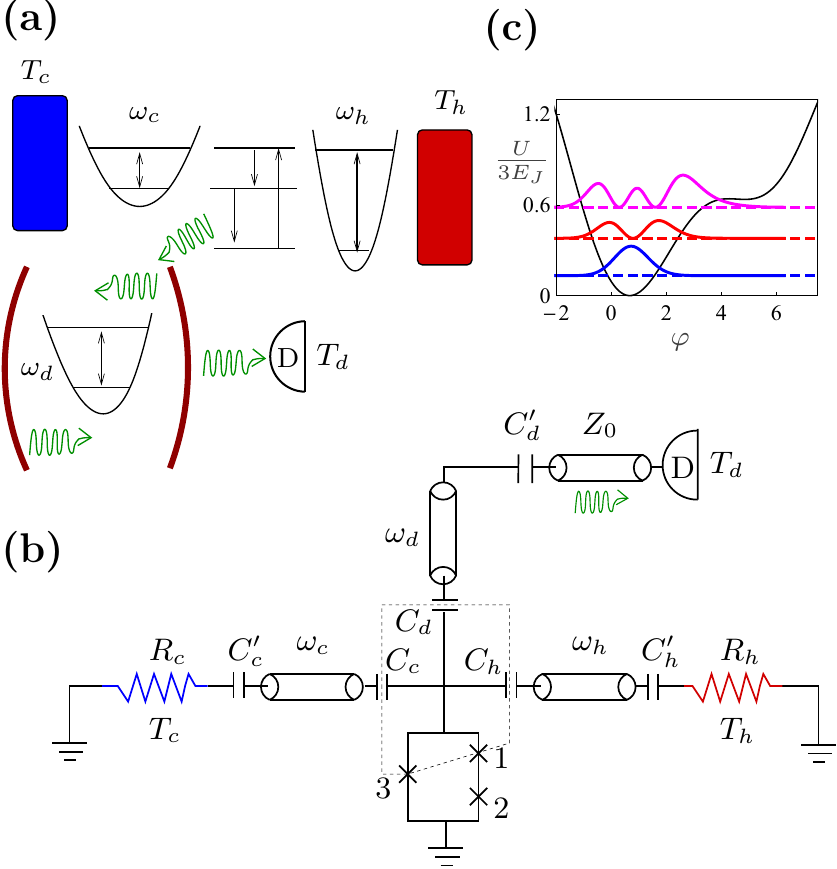}
\caption{(a) Schematics of a three level thermally driven maser.
Energy levels of the system are selectively coupled to the two reservoirs via filters.
(b) The circuit diagram of the proposed realization of the model shown in (a) with superconducting circuit components. 
(c) The potential of the loop  with three junctions at $\Phi/\Phi_0=0.32$ is shown by the solid black line. 
The horizontal dashed lines show the positions of the first three energy levels, 
while the  squares of the corresponding wavefunctions are  shown by solid curves.
}
 \label{model}
\end{center}
\end{figure}

The implementation of a heat driven maser, which allows conversion
of heat into coherent radiation,  has not yet been discussed in the cQED context.
At the same time, significant progress has been recently achieved  in
integrating the tools of ultrasensitive nanoscale bolometry into cQED devices \cite{Meschke,Timofeev,Schwab,Partanen,QHV, QHR, TPG2019}, 
and a unique platform for  studying the heat transport in the quantum limit has been created.  
Motivated by this progress, here we develop a theory and propose an experimental realization of a thermally driven single atom maser based on 
superconducting components.
We follow the original proposal by  Scovil and  Schulz-DuBois \cite{Scovil1959}, who have considered
a three level atom (qutrit), in which each interlevel transition is coupled to a separate thermal reservoir with
tunable temperature. Creating proper temperature differences between the reservoirs, one can induce population
inversion between the two energy levels coupled to the output port and in this way achieve lasing.
This device is an example of a quantum heat engine with the efficiency bound by the Carnot limit \cite{Scovil1967}.
We propose to use a superconducting loop with three Josephson junctions as an artificial atom with its three lowest
levels labeled as 0, 1, and 2. 
The artificial atom is capacitively coupled to the three high quality factor resonators which are tuned in resonance with
01, 12, and 02 transitions, see Fig. \ref{model}a. The hot (02) and cold (12) resonators are terminated by
resistors, which can be heated by bias currents, and the 01 resonator is connected to
the output port. The temperatures of the resistors can be monitored by normal metal - superconductor tunnel junction
thermometers \cite{Giazotto2009}. After proper calibration, the small variations of the resistor temperatures 
can be converted into power dissipated or emitted by them.
Our theory is based on the usual Lindblad equation formalism, which has been developed  
for a general laser setup by Mu and Savage \cite{Savage1992}, and has been later applied to the specific case
of thermally driven masers by Scully {\it et al} \cite{Scully2011} and by Li {\it et al}  \cite{Scully2017}.
We derive an explicit expression for the maser power in terms of circuit parameters and
show that it can reach up to few femtowatts for typical experimental conditions. 
This power is  several orders of magnitude higher than the lowest power detectable in experiment \cite{Meschke,Timofeev,Schwab,Partanen,
QHV, QHR, TPG2019, Kokkoniemi2019,Giazotto2020}.
We further show that one can detect the population inversion
between the states 0 and 1 without measuring the coherent radiation output of the device. 
Instead, one can connect a resistor to the output resonator and monitor how  the power dissipated in it varies with its temperature.
In presence of the population inversion the  power keeps flowing into the output resistor even if its temperature exceeds
the temperatures of the other two resistors. One can  understand this feature by assigning a negative effective temperature 
to the artificial atom with population inversion. In the heat transport context,  
negative temperature is hotter than any positive temperature, that is why the heat always flows into the resistor connected to the output resonator.

%\section{model}

\section{Model}
 
In our model, the artificial atom is realized as a superconducting loop 
with three Josephson junctions, similar to the conventional persistent-current
flux qubit \cite{Orlando, Devoret2017}, as shown in Fig. \ref{model} (b). 
The loop is formed by the two bridges connecting a superconducting island, which is
restricted by the capacitors $C_c,C_h,C_d$ and the junctions 1 and 3 and shown by dotted line in Fig. \ref{model} (b),
to the superconducting ground electrode. The left bridge contains only one Josephson junction
(junction 3), while the right bridge -- two other junctions (junctions 1 and 2).
The capacitors $C_c,C_h,C_d$ couple the island
to the cold, hot, and output (or detector) resonators respectively. The latter can be realized as
usual coplanar waveguide $\lambda/2$-resonators widely used in cQED circuits \cite{Esteve2008}.
The fundamental modes of the resonators have the frequencies $\omega_r$, where the index $r$ can be $h,c$ or $d$.
We denote the characteristic impedances of the resonators as $Z_{0r}$.
On the other side, the cold and the hot resonators are coupled to the resistors $R_c$
and $R_h$ via the capacitors $C'_c$ and $C'_h$.  
Finally, the capacitor $C'_d$ couples  
the output resonator to a resistor $R_d$ or to a transmission line with the impedance $Z_0$. 
The latter should guide the output radiation to the spectrum analyser. 
The temperatures $T_r$ of all resistors can be varied by applying DC bias currents to them without influencing the qutrit,
and can be monitored with normal metal - superconductor tunnel junction thermometers \cite{Giazotto2009}. 
The maser is pumped by heating up the hot resistor $R_h$.
The first three levels of the artificial atom have the energies $E_0$, $E_1$ and $E_2$, and the corresponding transition frequencies are
$\omega_{ij}=(E_i-E_j)/\hbar$ ($i,j= 0,1,2$).
These frequencies can be controlled by the magnetic flux $\Phi$ applied to the loop, and may be slightly detuned from the
frequencies of the corresponding resonators, so that $\omega_{h}=\omega_{20}+\Delta_h$, $\omega_{c}=\omega_{21}+\Delta_c$,
and $\omega_d=\omega_{10}+\Delta_d$. 
The asymmetry of the loop with three junctions makes the transitions between the states 0 and 2 possible.
Such transitions are required for the pumping of the maser.
The quality factors of the fundamental modes of the resonators \cite{Esteve2008}, $Q_r=\pi/(2\omega_r^2Z_{0r}R_r C_r^{\prime 2})$, 
should be sufficiently high to avoid their overlap with unintended transition frequencies,  $Q_r\gg\omega_r/|\omega_{10}-\omega_{21}|$.
However, $Q_r$ should not be too high to allow some tolerance in the detunings $\Delta_r$.

One can understand the operation principle of the maser first assuming  that the qutrit is 
uncoupled  from the output resonator and coupled only to the hot and the cold ones. The hot resistor induces
the transitions from the state $0$ to the state $2$ with the rate $\Gamma_{20}$, and the inverse transitions $2\to 0$ with the rate $\Gamma_{02}$.  
Likewise, the cold resistor is responsible for the transitions $1\to 2$ with the rate $\Gamma_{21}$ and $2\to 1$ with the rate $\Gamma_{12}$.
These rates satisfy the detailed balance relations $\Gamma_{02}/\Gamma_{20}=\exp(\hbar\omega_{20}/k_BT_h)$ 
and $\Gamma_{12}/\Gamma_{21}=\exp(\hbar\omega_{21}/k_BT_c)$ . 
One can easily verify that in this case the population inversion between the  
states 0 and 1 is achieved if \cite{Scovil1959} 
\begin{eqnarray}
\frac{\hbar\omega_{21}}{k_BT_c}\geq \frac{\hbar\omega_{20}}{k_BT_h}.
\label{condition1}
\end{eqnarray} 
Once the population inversion is there, one can weakly couple the output resonator to the qutrit.
The latter will be driven to a coherent state and will emit coherent radiation to the output transmission line.

\begin{figure*}
 \centering
 \includegraphics[scale=1]{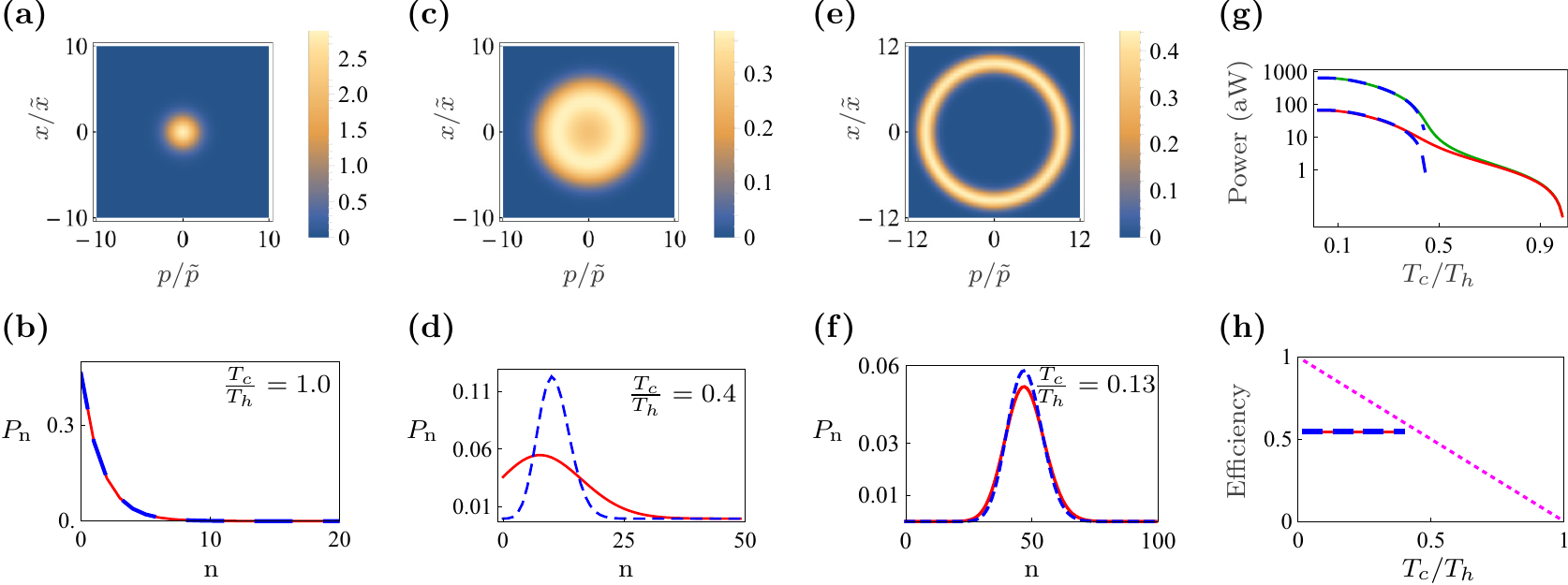}
\caption{(a), (c), (e)  Wigner functions of the output resonator, and (b), (d), (f) the corresponding photon number distributions (red solid lines).
The coordinate $x$ and the momentum $p$ of the oscillator modelling the output resonator are normalized 
with their zero point fluctuations $\tilde x$ and $\tilde p$.
The ratio  $T_c/T_h$  in the three sets takes the values 1.0, 0.4, 0.13, and $T_h=0.3$ K. 
The photon number distribution in (b) coincides with the Boltzmann distribution (blue dashed line). 
In (d) and (f) blue dashed lines show the Poissonian distribution.
We have used the following parameters: $E_C/2\pi\hbar=380$ MHz, $E_J/2\pi\hbar=5.2$ GHz, $C_c= 3$ fF, $C_h=18$ fF, $C_d=5 $ fF,
$R_c=80~\Omega$, $R_h=100~\Omega$, $C_{c}'=136$ fF, $C_{h}'=60$ fF, $C_{d}'=4$ fF,
$\omega_{h}/2 \pi=7.09~ {\rm GHz}$,
$\omega_{d}/2 \pi=3.87~ {\rm GHz}$,
$\omega_{c}/2 \pi=3.20~ {\rm GHz}$,
$Q_h\approx 44$, $Q_c\approx 53$, $Q_d\approx 66465$,
$\Phi/\Phi_0=0.32$, $\Delta_h=112$ MHz, $\Delta_c=-45$ MHz and $\Delta_d=24.6$ MHz.
(g) The lower red line shows the output power for $T_h=0.41$ K in a system with the parameters given above. 
The upper  green line shows the power (\ref{P_exact}) in a system with 
$C_c= 34$ fF, $C_h=76.5$ fF, $C_d=10 $ fF, $C_{c}'=289$ fF, $C_{h}'=127$ fF, $C_{d}'=4$ fF, 
$\Delta_h=1.2$ MHz, $\Delta_c=0.4$ MHz, $Q_h\approx 12$, $Q_c\approx 10 $ and remaining parameters are the same as before.
The approximation   (\ref{power}) is shown by the blue dashed line in both cases.
(h) Red line -- the efficiency found numerically, corresponding to the power represented with red line in (g), blue dashed line -- the ratio of frequencies ($\omega_{10}/\omega_{20}$),
dashed pink line -- Carnot efficiency.
}
\label{Pos_Wgn}
\end{figure*}

We now turn to the rigorous theoretical framework,
which allows one to evaluate maser power for given circuit parameters.  
We assume that all three junctions in the loop have the  same critical current $I_C$.
In this case, the Hamiltonian of the loop reads
\begin{eqnarray}
H_J &=& -4E_C\partial_\varphi^2 
+E_{J}\big[3\cos({\varphi_x}/{3})
\nonumber\\ &&
-\,\cos{\varphi}- 2\cos((\varphi-\varphi_x)/2)\big],
\label{H0}
\end{eqnarray}
where $\varphi$ is the Josephson phase difference between the central superconducting island of the device
and the ground, $E_C$ is the charging energy of the island, $\varphi_x=2\pi\Phi/\Phi_0$  
is the phase shift induced by the magnetic flux, $\Phi_0$ is the magnetic flux quantum and $ E_{J}=\hbar I_C/2e$ is the Josephson energy
of a single junction.
Here we follow Ref. \cite{Devoret2017} 
and assume that the phases of the junctions 1 and 2 are equal, $\varphi_1=\varphi_2=\varphi/2$.
The dependence of the potential on the phase difference $\theta=\varphi_1-\varphi_2$ may be
ignored if $E_J\gg E_C$, the self-capacitance of the island between the junctions is small, and one considers only
few low lying energy levels in the potential well. In this case, motion in $\theta$-direction should be frozen
because the resonators are coupled only to the phase $\varphi$.
Expanding the potential energy near the minimum $\varphi_x/3$, 
and keeping only the third and the fourth order terms in the phase difference $\bar\varphi=\varphi-\varphi_x/3$,
which is justified provided $3E_J\cos(\varphi_x/3)/2\gg E_C$, one obtains
the energy level spacings as 
$\hbar\omega_{10}=\sqrt{12 E_J E_C\cos(\varphi_x/3)}-3E_C/4$
and $\hbar\omega_{21}=\hbar\omega_{10}-3 E_C/4$ (see Appendix for details). 
The potential energy of the loop, numerically evaluated energy levels of the Hamiltonian (\ref{H0}) and the square of the corresponding wave functions  
are shown in Fig. \ref{model}(c). 
  
Considering only the three lowest levels of the qutrit and high quality factor resonators, one can express the Hamiltonian of the system in the form
\begin{eqnarray}
&& H = \sum_{n=0}^2 E_n|n\rangle\langle n|+ \sum_{r=h,c,d}\hbar \omega_r\bigg(a^{\dagger}_r a_r +{1}/{2}\bigg)
\nonumber\\ &&
+\, \hbar g_d[(a^{\dagger}_d +a_d)(\sigma^+_{01}  + \sigma^-_{01})]
+ \hbar g_h[(a^{\dagger}_h +a_h)(\sigma^+_{02}  + \sigma^-_{02})]
\nonumber\\ &&
+\, \hbar g_c[(a^{\dagger}_c +a_c)(\sigma^+_{12}  + \sigma^-_{12})].
\label{H}
\end{eqnarray}
Here $g_r$ are the coupling constants between the qutrit and the resonators  given by 
\begin{eqnarray}
g_{mn}=\frac{\omega_r^2C_r}{2e}\sqrt{\frac{\hbar Z_{0r}}{\pi}}\varphi_{mn},
\label{g_mn}
\end{eqnarray}
with $g_h=g_{20}$, $g_c=g_{21}$ and $g_d=g_{10}$ (for explicit approximate expressions see Eqs. (\ref{gd})),   
$\varphi_{ij}=\langle i| \hat\varphi |j\rangle$ 
are the matrix elements of the Josephson phase between the states of the qutrit,  
$a_r$ and $a^{\dagger}_r$ are the ladder operators of the resonators, and
$\sigma_{ij}^{+}=|i \rangle\langle j|$
and $\sigma^-_{ij}=|j\rangle\langle i|$ are the transition operators between the levels of the qutrit.
The damping rates of the fundamental modes of the resonators, $\kappa_r=(2/\pi)\omega_r^3 Z_{0r}R_rC_{r}^{\prime 2}$, 
are induced by their coupling to the resistors $R_r$, or, in case of the output resonator -- to the outer transmission line with the impedance $Z_0$,
in that case, one should replace $R_d$ by $Z_0$.
Below we assume $\kappa_h/g_h\gtrsim 1$ and $\kappa_c/g_c\gtrsim 1$ and ignore 
coherent coupling between the qutrit and the hot and cold resonators, 
treating the latter as parts of the environment inducing the transitions between the qutrit levels. 
However, at this stage we keep the ratio $\kappa_d/g_d$ arbitrary. In this case, 
the  Lindblad equation governing the time evolution of the density matrix $\rho$ of the "qutrit plus output resonator" system has the form
\begin{eqnarray}
 \frac{d \rho}{dt}&=&-\frac{i}{\hbar}[H_{q-d},\rho]+\kappa_d (N_d^{\rm eq}+1)\bigg(a\rho a^{\dagger}-\frac{1}{2}\{a^{\dagger}a,\rho\}\bigg)
\nonumber\\ &&
 +\,\kappa_d N_d^{\rm eq}\bigg(a^{\dagger}\rho a-\frac{1}{2}\{a a^{\dagger},\rho\}\bigg)
\nonumber\\ &&
+\,\sum_{i=1}^2\sum_{j=0}^{i-1} \Gamma_{ij}\bigg(\sigma_{ij}^{-}\rho \sigma_{ij}^{+}
-\frac{1}{2}\{\sigma_{ij}^{+}\sigma_{ij}^{-},\rho\}\bigg)
\nonumber\\ &&
+\,\sum_{j=1}^2\sum_{i=0}^{j-1}\Gamma_{ij}\bigg(\sigma_{ij}^{+}\rho \sigma_{ij}^{-}-\frac{1}{2}\{\sigma_{ij}^{-} \sigma_{ij}^{+},\rho\}\bigg).
\label{mastereq}
\end{eqnarray}
Here  $H_{q-d}$ is the Hamiltonian describing "qutrit plus output resonator" system having the form (\ref{H}) with $g_h=g_c=0$, and $r=d$,
$N_d^{\rm eq}=1/(e^{\hbar\omega_d/k_BT_d}-1)$ is the equilibrium photon number in the output resonator, 
\begin{eqnarray}
\Gamma_{ij} &=& |\varphi_{ji}|^2 \frac{\hbar\omega_{ji}}{2 e^2}
\,{\rm Re}\left[\frac{1+n_h(\omega_{ji})}{Z_h(\omega_{ji})}+\frac{1+n_c(\omega_{ji})}{Z_c(\omega_{ji})}\right],\; (i<j)
\nonumber\\
\Gamma_{ij} &=& |\varphi_{ij}|^2 \frac{\hbar \omega_{ij}}{2 e^2}
\,{\rm Re}\left[\frac{n_h(\omega_{ij})}{Z_h(\omega_{ij})}+\frac{n_c(\omega_{ij})}{Z_c(\omega_{ij})}\right],\; (i>j)
\label{Gamma_mn}
\end{eqnarray}
are the transition rates from the state $j$ into the state $i$,  
$Z_h(\omega),Z_c(\omega)$ are the impedances of the hot and cold resonators (\ref{Zj}), 
and $n_{h}(\omega)$ and $n_{c}(\omega)$ are the effective photon distributions 
in these resonators. According to Eqs. (\ref{Gamma_mn}) both hot and cold resonators
contribute to all transitions, thus the expressions (\ref{Gamma_mn}) are valid
even for low quality resonators with overlapping spectrum lines.
In the limit $\Gamma_{02}\ll\kappa_h$ and $\Gamma_{12}\ll\kappa_c$ the resonators
are almost decoupled from the qutrit and $n_{h,c}(\omega)=1/(e^{\hbar\omega/k_BT_{h,c}}-1)$.
For $\Gamma_{02}\gtrsim \kappa_{h}$, $\Gamma_{12}\gtrsim \kappa_{c}$ the distributions $n_{h,c}(\omega)$ should be found self consistently 
from the equation of motion for the average values of the ladder operators $\langle a_{h,c}\rangle$.  
For high quality factor resonators, with non-overlapping spectral lines, and in the semiclassical approximation, 
the equations for $\langle a_{h,c}\rangle$ can be re-written
in terms of the average numbers of photons in the resonators $N_h = \langle a^\dagger_h a_h\rangle\approx n_h(\omega_{20})$ and 
$N_c=\langle a^\dagger_c a_c\rangle\approx n_c(\omega_{21})$, see  Eqs. (\ref{dot_Nh},\ref{dot_Nc}).
In the experimentally relevant limit $3E_J\cos(\varphi_x/3)/2\gg E_C$ 
the phase matrix elements can be approximately evaluated by means of the perturbation theory in $\bar\varphi$. In this way, one finds
$|\varphi_{10}|^2 = \sqrt{4E_C/3E_J\cos(\varphi_x/3)}$, $|\varphi_{21}|^2 = 2|\varphi_{10}|^2$ 
and $|\varphi_{20}|^2=(E_C/54E_J)\tan^2(\varphi_x/3)/\cos(\varphi_x/3)$
(see Sec. \ref{Perturbation} of the Appendix for details). 
We note that the matrix element $\varphi_{20}$, needed for pumping the maser, 
differs from zero only in presence of magnetic field.
In the numerical simulations we compute the matrix elements exactly instead of using these approximate expressions.
We use the parameters such that the third energy level is close to the barrier top of the double-well potential profile, as shown in Fig. \ref{model} (c). This
increases the anharmonicity $\omega_{10}-\omega_{21}$ above the value $3E_C/4\hbar$, predicted by the perturbation theory in $\bar\varphi$,
and helps to increase the maser power.

Instead of numerically solving the Lindblad equation (\ref{mastereq}),
for most practical purposes one can use a simpler semiclassical model of the device 
valid for high quality factor resonators without overlap of the spectral lines 
and for high numbers of photons in the output resonator $N_d=\langle a^\dagger_d a_d\rangle\gtrsim 1$.
Similar models are often used in the literature,
see e.g. Refs. \cite{Savage1992,Scully2011,Scully2017}.
In this approximation, the system dynamics is governed by master equation for the vector of
occupation probabilities of the qutrit states ${\bm p}=(p_0,p_1,p_2)^T$,
\begin{eqnarray}
\dot{\bm p}={\bm \Gamma}{\bm p}
\label{master}
\end{eqnarray}
with the transition rates matrix
\begin{equation}
{\bm \Gamma}=\left[ {\begin{array}{ccc}
  -\Gamma_{10}-\Gamma_{20}&\Gamma_{01}&\Gamma_{02}\\
   \Gamma_{10}&-\Gamma_{01}-\Gamma_{21}&\Gamma_{12}\\
   \Gamma_{20}&\Gamma_{21}&-\Gamma_{02}-\Gamma_{12}\\
  \end{array} 
  } \right],
\label{rates_matrix}
\end{equation}
and by the equations for the numbers of photons,
\begin{eqnarray}
\dot N_d &=& \Gamma_{01}p_1 - \Gamma_{10}p_0 - \kappa_d (N_d - N_d^{\rm eq}),
\label{dot_Nd}\\
\dot N_h &=& \Gamma_{02}p_2 - \Gamma_{20}p_0 - \kappa_h (N_h - N_h^{\rm eq}),
\label{dot_Nh}\\
\dot N_c &=& \Gamma_{12}p_2 - \Gamma_{21}p_1 - \kappa_c (N_c - N_c^{\rm eq}).
\label{dot_Nc}
\end{eqnarray}
Here $N_r^{\rm eq}=1/(e^{\hbar\omega_r/k_BT_r}-1)$ are the equilibrium photon numbers.
Since the resonator lines do not overlap, the transition rates (\ref{Gamma_mn}) take the simple form
\begin{eqnarray}
\Gamma_{01} = \gamma_d (1+N_d),\;\;\Gamma_{10} =\gamma_d N_d,
\label{Gamma_01}\\
\Gamma_{02} = \gamma_h (1+N_h),\;\;\Gamma_{20} =\gamma_h N_h,
\label{Gamma_02}\\
\Gamma_{12} = \gamma_c (1+N_c),\;\;\Gamma_{21} =\gamma_c N_c,
\label{Gamma_12}
\end{eqnarray} 
where the rates of downward transitions in the absence of photons in the resonators are
\begin{eqnarray}
&& \gamma_h =  \frac{\hbar \omega_{20}}{2 e^2}\,{\rm Re}\left[\frac{|\varphi_{20}|^2}{Z_h(\omega_{20})}\right],\;
\gamma_c =  \frac{\hbar \omega_{21}}{2 e^2}\,{\rm Re}\left[\frac{|\varphi_{21}|^2}{Z_c(\omega_{21})}\right],
\nonumber\\
&& \gamma_d =  \frac{\hbar \omega_{10}}{2 e^2}\,{\rm Re}\left[\frac{|\varphi_{10}|^2}{Z_d(\omega_{10})}\right].
\label{gamma_hc}
\end{eqnarray}
These rates can be simplified for small detunings $\Delta_r$, 
\begin{eqnarray}
\gamma_r = \frac{g_r^2\kappa_r}{\Delta_r^2 + \kappa_r^2/4},\;\; r=h,c,\;{\rm or}\, d.
\label{gamma_r}
\end{eqnarray}
Eqs. (\ref{master}) and (\ref{dot_Nd}-\ref{dot_Nc}) can be easily solved numerically in the stationary case.
Once the photon numbers $N_r$ are known, all other parameters can  be evaluated. 
For example, the output power is given by the  expression (see Appendix \ref{Master} for details):  
\begin{eqnarray}
\frac{P_{out}}{\hbar\omega_d} = \frac{\gamma_h\gamma_c\gamma_d[(1+N_d)N_h(1+N_c)-N_d(1+N_h)N_c]}{B},
\nonumber\\
\label{P_exact}
\end{eqnarray}
where
\begin{eqnarray}
&& B = \gamma_d\gamma_h[(1+2N_h)(1+N_d)+(1+N_h)N_d] 
\nonumber\\ &&
+\, \gamma_d\gamma_c[(1+N_c)(1+N_d)+(1+2N_c)N_d] 
\nonumber\\ &&
+\, \gamma_h\gamma_c[N_h (1+N_c)+(1+N_h)N_c+N_hN_c].
\end{eqnarray}

\section{Results and discussion}

We numerically find the stationary solution of Eq. (\ref{mastereq}) in the standard way \cite{Savage1992}.
In Figs.  \ref{Pos_Wgn}a- \ref{Pos_Wgn}f 
we plot the resulting photon number distributions and Wigner functions of the output resonator  
for several values of the temperature bias and assuming typical circuit parameters given in the caption.
These parameters result in the following values of
the coupling constants and the damping rates for hot and cold resonators:
$g_h/2 \pi\approx45.34~{\rm MHz} < k_h/2 \pi \approx 161.17$ MHz  and
$g_c/2 \pi \approx 22.68 ~{\rm MHz}< k_c/2 \pi \approx 60.90$ MHz. For the  output resonator,  we obtain 
$g_d/2 \pi\approx 29.88 ~{\rm MHz}\gg k_d/2 \pi\approx 58.21$ KHz. We also obtain $\gamma_h=52.25$ MHz, $\gamma_c=33.94$ MHz,
and $\gamma_d=52.88$ MHz.
Although for these parameters, the ratios $\gamma_h/\kappa_h$ and $\gamma_c/\kappa_c$ are small, 
they are not small enough to assume equilibrium photon numbers in the hot and cold resonators. That is why  
we have used Eqs. (\ref{dot_Nh},\ref{dot_Nc}) in oder to find the stationary values of $N_h$ and $N_c$.
We observe that the photon number distribution evolves from the Boltzmann form $e^{-\hbar\omega_dn/k_BT_d}$ for $T_d=T_h=T_c$ (Fig. \ref{Pos_Wgn}b) to
the Poissonian form $ P^{\rm Pois}_n=e^{-\langle n\rangle} \langle n\rangle^n/n!$ typical for a lasing state (Fig. \ref{Pos_Wgn}f)
as the temperature bias becomes stronger. Simultaneously, the Wigner function evolves from a gaussian thermal distribution (Fig. \ref{Pos_Wgn}a)
to a ring-shaped form indicating high amplitude harmonic oscillations in the coherent state (Fig. \ref{Pos_Wgn}e). 

Following the recepie of Refs. \cite{Scully2011,Scully2017}, 
we derive an approximate expression 
for the output power of the maser in the lasing regime $N_d\gg 1$, $N_h\gtrsim N_c$ and $\gamma_d\gg\kappa_d(1+3N_c)$.
Simplifying the exact expression (\ref{P_exact}) in this limit we obtain
\begin{eqnarray}
P_{out}=\frac{\gamma_c\gamma_h(N_h-N_c)}{\gamma_c(2+3 N_c)+\gamma_h(2+3 N_h)}\hbar \omega_{10}.
\label{power}
\end{eqnarray}
Eq. (\ref{power}) rather  well agrees with the exact numerics for the chosen parameter values, see Fig. \ref{Pos_Wgn} (g).
It provides a good approximation in the high temperature bias regime $T_c/T_h\ll \omega_{10}/\omega_{20}$, in which lasing
is expected. In Fig. \ref{Pos_Wgn} (g) we also plot the exact power values [Eq. (\ref{P_exact})] with red line, which in high 
temperature bias regime exactly matches with the power obtained by solving the Lindblad equation (\ref{mastereq}), while
the green line represents high power regime  [Eq. (\ref{P_exact})] for a certain choice of parameters.
In  the plot for high power (green line), we use  $g_h/2 \pi=191.73$ MHz, $g_c/2 \pi=258.16$ MHz, $g_d/2 \pi=74.71$ MHz,
$\kappa_h/2 \pi=577.86$ MHz, $\kappa_c/2 \pi=311.46$ MHz, and $\kappa_d/2 \pi=58.22$ KHz.

For typical system parameters $\omega_{10}/2\pi \approx 5$ GHz and $E_C/2\pi\hbar\approx500$ MHz, $T_h\approx 0.35$ K, and
$T_c\lesssim 0.05$ K,
the maser operates in the regime $N_h\sim 1$ and $N_c\ll 1$.
For sufficiently strong coupling to the hot resonator, $\gamma_h\gg \gamma_c$,
the power (\ref{power}) approaches the limiting value $\tilde P_{out}=\hbar\omega_{10}\gamma_c/3$. Increasing the rate $\gamma_c$
by design, one can push this power up. However, in our model $\gamma_c$ cannot exceed $\kappa_c$,
otherwise the cold resonator cannot be efficiently cooled. The damping rate $\kappa_c$, in turn, is limited by the condition
$\kappa_c\lesssim (\omega_{10}-\omega_{21})/2 = 3E_C/8$. Indeed, for higher $\kappa_c$  
the output resonator becomes coupled to the $1\leftrightarrow 2$ transition and gets cooled. 
Thus, at most, one can achieve $\gamma_c\sim 3E_C/8$ 
and the maximum output power of the device can be estimated as $P_{\max}\approx \omega_{10}E_C/8$.
For typical device parameters given above one finds $P_{\max}=1.3$ fW. 
This value is comparable with the maser power of 0.7 fW reported in Ref. \cite{Astafiev2007}, 
but it is significantly lower than 255 fW observed in a laser pumped by a voltage biased Josephson junction \cite{Cassidy}. 
The linewidth of the maser can be estimated 
according to Schawlow-Townes formula \cite{SiegmanBook}, $\delta \omega=\kappa/ 2N_d$. For maximum maser power one finds
$N_d=P_{\max}/\hbar\omega_d\kappa_d\approx E_C/8\hbar\kappa_d$ and 
the linewidth reduces to $\delta \omega\approx4\hbar\kappa_d^2/ E_C= 4\hbar \omega_d^2/E_C Q_d^2$.
For the parameters quoted above and for $Q_d\sim 10^3$ we estimate $\delta\omega/2\pi\sim 0.2$ MHz.
The efficiency of the maser is defined as the ratio of the output maser power to the input heat absorbed
from the hot resistor, $\eta=P_{out}/P_{in}$. One can show \cite{Scovil1959} that in the lasing regime  
$\eta\approx\omega_{10}/ \omega_{20}$. This value, however, cannot exceed the Carnot efficiency
$1-T_c/T_h$. Indeed, for $\omega_{10}/ \omega_{20} > 1-T_c/T_h$, no population inversion in the qutrit
can be created (see Eq. (\ref{condition1})) and, hence, no lasing can occur. Instead, in this regime, the trivial heat transfer from the hot resistor
to the output port is going on. The efficiency of the maser is plotted in Fig. \ref{Pos_Wgn} (h).
Finally, the lasing threshold is determined by the condition (\ref{threshold}),
which can be satisfied only if $\kappa_d\ll \gamma_d$.

 \begin{figure}
\begin{center}
\includegraphics [width=8cm] {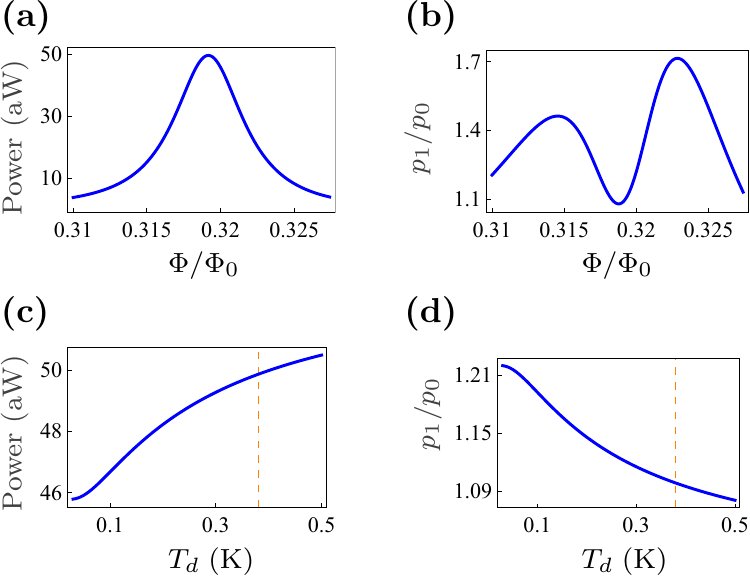}
\caption{(a) Power dissipated in the detector resistor versus applied magnetic flux; (b) the ratio $p_1/p_0$ as a function of external magnetic flux;
(c)  power dissipated in the detector resistor versus the detector temperature $T_d$ (blue line) and $T_d=T_h$ (red dashed line); 
(d) $p_1/p_0$ versus $T_d$ (blue line) and $T_d=T_h$ (red dashed line).
In (a) and (b), the temperature of the 
the detector is fixed at $T_d= 0.03$ K. 
In (c) and (d), $\Phi/\Phi_0=0.32$, and $T_d$  varies from 0.03 K to 0.5 K. The direction of heat flow does not change
even when $T_d> T_h$. 
We have used the following parameter values: $C_d= 1$ fF, $C_{d}'=100$ fF, $R_d=35$ $\Omega$,
$R_c=130$ $\Omega$, $C_{c}=4$ fF, $R_h=110$ $\Omega$, $\omega_h/ 2\pi=7.06$ GHz, $\omega_c/ 2\pi=3.21$ GHz,
$\omega_d/ 2\pi=3.87$ GHz,
$T_c=0.03$ K and $T_h=0.38$ K. Other parameters are same as in the caption of Fig. \ref{Pos_Wgn} (e).
The quality factors of the resonators are $Q_d\approx 152$,
 $Q_c\approx32$, and  $Q_h\approx 40$. 
 In numerical simulation, we have kept six  lowest levels of the artificial atom and assumed equilibrium photon numbers in all the resonators
 for simplicity and considered all the possible transitions due to all the three resonators. 
The maximum population of the sixth level, achieved at $T_d=0.5$ K,  is found to be $p_5\approx0.05$.
}
\label{PoPo}
\end{center}
\end{figure}

Ideally, the output radiation of the maser should be detected with a spectrum analyser.
We will now demonstrate that one can replace the latter by the simplest broadband detector -- a resistor $R_d$ --
and still detect the population inversion in the
qutrit. Let us  assume that the coupling between all three resonators and the resistors is 
stronger than the coupling between them and the qutrit,  $C_{r}'\gg C_r$. No lasing can occur in this regime, and the photons in all
resonators are equilibrated with the corresponding resistors, $N_r=N_r^{\rm eq}$. 
The output power of the device (\ref{P_exact})
is proportional to the combination $P_{out}\propto (1+N_d^{\rm eq})N_h^{\rm eq}(1+N_c^{\rm eq}) - N_d^{\rm eq}(1+N_h^{\rm eq})N_c^{\rm eq}$. It
is the net contribution of the forward process, in which one photon comes from the hot resonator
and two photons are subsequently emitted by the qutrit into cold and output resonators, and of the reverse process.  
Substituting Bose functions in the above expression, one can check that  $P_{out}>0$  if and only if
\begin{eqnarray}
\frac{\hbar\omega_d}{k_BT_d} + \frac{\hbar\omega_c}{k_BT_c} > \frac{\hbar\omega_h}{k_BT_h}.
\label{condition}
\end{eqnarray}
Equation\ref{condition} ensures that the entropy production for the net forward process is positive.
As we have discussed above, for $g_d\ll g_c,g_h$ the population inversion
in the qutrit appears if the condition (\ref{condition1}) is satisfied. Obviously, in this case
the inequality (\ref{condition}) is satisfied, and hence $P_{out}>0$, for any temperature of the output resonator $T_d$
including high temperatures $T_d>T_h$. This unusual behaviour discussed by Geusic \textit{et al} \cite{Scovil1967} is illustrated in the context 
of superconducting circuits in Fig. \ref{PoPo} for a certain choice of the system parameters.
In Figs. \ref{PoPo}a,c we plot the power dissipated in the resistor $R_d$ and 
the ratio of the occupation probabilities  $p_1/p_0$ for a fixed 
temperature $T_d=0.03$ K as functions of the magnetic flux. 
In Figs. \ref{PoPo}c and \ref{PoPo}d, we take the value of the magnetic flux $\Phi/\Phi_0=0.32$ close to the maximum power, 
and  plot this power together with the ratio $p_1/p_0$ versus the temperature $T_d$.
$P_{out}$ indeed remains positive even for $T_d>T_h=0.38$ K.
Thus, we have shown that in presence of the population inversion between the levels 0 and 1 heat can flow from the colder resistor 
$R_h$ to the hotter resistor $R_d$. One can also interpret this observation as follows: population inversion
implies negative temperature of the qutrit, and the latter is formally hotter than any positive temperature.
Clearly, this property also holds in the lasing regime where the photon number distribution in the output resonator strongly 
deviates from equilibrium.

\section{Conclusion}

In conclusion, we have proposed a model of thermally driven on-chip three level maser
in  a superconducting circuit,  which converts heat into coherent radiation.
We have also proposed an experimentally feasible method of detecting
the population inversion in the artificial atom by thermometry.
We have derived a simple analytical expression for the output power of the maser in terms of the circuit parameters. 
We have shown that powers up to few femtowatts can be achieved in a typical cQED setup.
We believe that the proposed way of converting heat produced by DC-biased resistors into coherent radiation on-chip 
can be very useful for quantum circuit applications. The same technique can be used to control and detect the populations of
energy levels of quantum superconducting devices.

Our model can be generalized to a heat accelerator and a heat switch setups if one allows 
manipulating the coupling between the qutrit and one of the heat baths. 
One can also include more than one qutrit into it
and consider more subtle effects like, for example, 
superradiance of several artificial atoms \cite{Ke2019}.
Finally, our analysis can be generalized to a four level laser model,
which is, however, more challenging to realize in the experiment.

\section{Acknowledgement}
 
This work was supported by the Academy of Finland Centre of Excellence
program (project No. 312057), European Union's Horizon 2020 research 
and innovation programme under the European Research Council (ERC) 
programme (grant agreement No. 742559) and Marie Sklodowska-Curie grant agreement No. 843706. 
We acknowledge the computational resources provided by the Aalto Science-IT project.

\appendix

\section{Approximate theory of the qutrit}
\label{Perturbation}

We consider the setup shown in Fig. \ref{model}b and assume that all Josepshon junctions in the superconducting loop have the same critical current  $I_C$.
In this case, the phase difference on the junction 3 equals to $\varphi$ and on the junctions 1 and 2 it is $\varphi'/2$.
We can assume that the phase drops across junction 1 and 2 are equal when 
 $E_J\gg E_{C}$ 
and capacitance to the ground ($C_0$) by the island between the junctions 1 and 2 is $C_0\ll C_J/4$, where $C_J$ is the capcitance of a junction \cite{Devoret2017}.
The total current through the loop is $I= I_{C}\sin{(\varphi'/2)}+I_{C}\sin\varphi$.
The flux quantization condition implies $\varphi_3=\varphi '+2\pi  \Phi/\Phi_0$. This leads to
\begin{equation}
 I= I_{C}[\sin{(\varphi-\varphi_x)/2}+\sin{\varphi}],
\label{current2}
\end{equation}
where $\varphi_x=2\pi\Phi/\Phi_0$. The potential energy of the loop reads
\begin{eqnarray}
 U=\frac{\hbar}{2e}\int_{\varphi_x/3}^\varphi I_{C}(\sin{(\varphi_1-\varphi_x)/2}+\sin{\varphi_1}) d\varphi_1\nonumber\\
 =E_J[3\cos{(\varphi_x/3)}-\cos{\varphi}-2\cos{((\varphi-\varphi_x)/2)}].
 \label{Uenergy}
\end{eqnarray}
Combining this expression with the charging energy of the island $-4E_C\partial^2_\varphi$, we arrive at the
Hamiltonian of the artificial atom (\ref{H0}).

The potential energy (\ref{Uenergy}) achieves the minimum value $U=0$
at $\varphi=\varphi_x/3$ provided $-\pi\leq\varphi_x\le\pi$. 
Expanding it near the minimum 
up to the quartic order in $\bar{\varphi}=\varphi-\varphi_x/3$, 
we approximate the Hamiltonian (\ref{H0}) as 
\begin{equation}
 H=-4E_C\frac{\partial^2}{\partial\bar{\varphi}^2}+\tilde E_J\left[\frac{\bar{\varphi}^2}{2}
-\tan\frac{\varphi_x}{3}\frac{\bar{\varphi}^3}{12}-\frac{\bar{\varphi}^4}{32}\right],
 \label{H_total}
\end{equation}
where  $\tilde E_J=3E_{J}\cos{(\varphi_x/3)}/2$.
In the lowest order perturbation theory in $\propto\bar\varphi^3,\bar\varphi^4$ we obtain
the usual expression for the low lying energy levels of a quarctic oscillator
  \begin{eqnarray}
 E_n=\sqrt{8 \tilde E_J E_C}\left(n+\frac{1}{2}\right)-\frac{E_C}{16}(6n^2+6n+3),
 \end{eqnarray}
which is valid provided $E_C\ll\tilde E_J$. The correction $\propto\bar\varphi^3$ does not affect
the energy levels due to symmetry.
 The transition frequencies between the three lowest levels are
\begin{eqnarray}
\hbar\omega_{10} &=& E_1-E_0=(8 \tilde E_J E_C)^{1/2}-{3E_C}/{4},
\nonumber\\ 
\hbar\omega_{21} &=& E_1-E_0 = (8 \tilde E_J E_C)^{1/2}-3E_C/2,
\nonumber\\
\hbar\omega_{20} &=& E_2-E_0 =2(8 \tilde E_J E_C)^{1/2}-{9E_C}/{4}.
\end{eqnarray}

The corrected $n$th eigenstate is given by
$|n\rangle=|n\rangle_2+|n\rangle_3 +|n\rangle_4$, where $|n\rangle_2$ is
the eigenstate of the unperturbed quadratic Hamiltonian,
the correction coming from $\propto\bar{\varphi}^3$ term is denoted as $|n\rangle_3$ and the
corrections due to $\propto\bar{\varphi}^4$ term --- as $|n\rangle_4$.
These corrections read
 \begin{eqnarray}
 |n\rangle_3&=&-\frac{1}{24}\left(\frac{4E_C}{3 E_J}\right)^{{1}/{4}}\frac{\tan({\varphi_x}/{3})}{\left(\cos({\varphi_x}/3)\right)^{{1}/{4}}}
\nonumber\\ &&\times\,
 \big[\big(\sqrt{n(n-1)(n-2)}/3\big)|n-3\rangle_2\nonumber\\
&&-\big(\sqrt{(n+1)(n+2)(n+3)}/3\big) |n+3\rangle_2\nonumber\\
  &&+3n\sqrt{n}|n-1\rangle_2 \nonumber
 -3(n+1)\sqrt{(n+1)}|n+1\rangle_2\big],
\end{eqnarray}
\begin{eqnarray}
 |n\rangle_4 &=& -\frac{1}{64\sqrt{2}}\sqrt{\frac{E_C}{E_J\cos\frac{\varphi_x}{3}}}  
\Biggl[(4n-2)\sqrt{n(n-1)}|n-2\rangle_2 \nonumber\\
 &&- (4n+6)\sqrt{(n+1)(n+2)} |n+2\rangle_2\nonumber\\
  &&-(1/2)\sqrt{(n+1)(n+2)(n+3)(n+4)}|n+4\rangle_2 \nonumber\\
&& +(1/2)\sqrt{n(n-1)(n-2)(n-3)}|n-4\rangle_2\Biggl].
\end{eqnarray}
Having found the corrections to the eigenfunctions, we can evaluate
the phase matrix elements:
\begin{eqnarray}
&& \langle n-1| \hat\varphi |n\rangle
= \left(\frac{4 E_C}{3 E_J\cos\frac{\varphi_x}{3}}\right)^{1/4}\sqrt{n}\left[ 1+n\sqrt{\frac{E_C}{32E_J}} \right],
\nonumber\\ &&
\langle n-2| \hat\varphi |n\rangle
=\sqrt{\frac{E_C}{3 E_J\cos\frac{\varphi_x}{3}}}\frac{\tan\frac{\varphi_x}{3}}{6}\sqrt{ n(n-1)},
\nonumber\\ &&
\langle n-3| \hat\varphi |n\rangle
=\, \left(\frac{4 E_C}{3 E_J\cos\frac{\varphi_x}{3}}\right)^{3/4}\frac{\sqrt{n(n-1)(n-2)}}{48}.
\label{phinm3}
\end{eqnarray}

With the explicit expressions for the matrix elements (\ref{phinm3}), we find the approximate expressions for the
coupling constants (\ref{g_mn}) between the qutrit and the resonators,
\begin{eqnarray}
g_h &=& \frac{\omega_h^2C_h}{2e}\sqrt{\frac{\hbar Z_{0h}}{\pi}}\sqrt{\frac{2E_C}{3 E_J\cos\frac{\varphi_x}{3}}}\frac{\tan\frac{\varphi_x}{3}}{6},
\nonumber\\ 
g_c &=&\sqrt{2}\, \frac{\omega_c^2C_c}{2e}\sqrt{\frac{\hbar Z_{0c}}{\pi}}\left(\frac{4E_C}{3E_J\cos\frac{\varphi_x}{3}}\right)^{1/4},
\nonumber\\ 
g_d &=& \frac{\omega_d^2C_d}{2e}\sqrt{\frac{\hbar Z_{0d}}{\pi}}\left(\frac{4E_C}{3E_J\cos\frac{\varphi_x}{3}}\right)^{1/4}.
\label{gd}
\end{eqnarray}

\section{Impedances of the resonators}

The transition rates between the states of the qutrit (\ref{Gamma_mn}) are expressed in terms
of the impedances of the resonators. These impedances include the resonators themselves as well as
the coupling capacitors and the resistors, which terminate the resonators and serve as thermal baths
in our setup. The impedance of the resonator $r$ (where $r=h,c$ or $d$) reads
\begin{eqnarray}
Z_r(\omega)&=&Z_{0r}\frac{\left(R_r + \frac{1}{-i\omega C'_r}\right)\cos\omega t_r -i Z_{0r}\sin\omega t_r}
{Z_{0r} \cos\omega t_r -i \left(R_r + \frac{1}{-i\omega C'_r}\right)\sin\omega t_r}
\nonumber\\&&
+\,\frac{1}{-i \omega C_r}.
\label{Zj}
\end{eqnarray}
Here $t_r$ are the travel times of the microwaves along the resonators.

\section{Maser output power and lasing threshold}
\label{Master}

Solving the system (\ref{master}), we find the steady state occupation probabilities of the qutrit states in the form
\begin{eqnarray}
&& p_0=\frac{\Gamma_{01} +  \frac{\Gamma_{02}\Gamma_{21}}{\Gamma_{02}+\Gamma_{12}}}{A},\;\;
p_1 = \frac{\Gamma_{10} + \frac{\Gamma_{12}\Gamma_{20}}{\Gamma_{02}+\Gamma_{12}}}{A},
\label{p01}\\ &&
 A=\left(1+\frac{\Gamma_{20}}{\Gamma_{02}+\Gamma_{12}}\right)\left(\Gamma_{01} +  \frac{\Gamma_{02}\Gamma_{21}}{\Gamma_{02}+\Gamma_{12}} \right)
\nonumber\\ &&
+\, \left(1+\frac{\Gamma_{21}}{\Gamma_{02}+\Gamma_{12}}\right)\left(\Gamma_{10} + \frac{\Gamma_{12}\Gamma_{20}}{\Gamma_{02}+\Gamma_{12}}\right),
\nonumber\\ &&
p_2 = 1-p_0-p_1.
\end{eqnarray}
The steady state numbers of photons in the resonators follow from Eqs. (\ref{dot_Nd}),
\begin{eqnarray}
N_d &=& N_d^{\rm eq}+(\Gamma_{01}p_1-\Gamma_{10}p_0)/\kappa_d,
\label{Nd}\\
N_h &=& N_h^{\rm eq}+(\Gamma_{02}p_2-\Gamma_{20}p_0)/\kappa_h,
\label{Nh}\\
N_c &=& N_c^{\rm eq}+(\Gamma_{12}p_2-\Gamma_{21}p_1)/\kappa_c.
\label{Nc}
\end{eqnarray}
Since the rates $\Gamma_{ij}$ (\ref{Gamma_01}-\ref{Gamma_12}) depend on the photon numbers $N_r$, 
these equations should be numerically solved for $N_r$. 

The output power of the maser reads
\begin{eqnarray}
P_{out} =  \hbar\omega_d(\Gamma_{01}p_1-\Gamma_{10}p_0).
\label{P1}
\end{eqnarray}
With the aid of the explicit solutions (\ref{p01}) and the expressions for the rates (\ref{Gamma_01}-\ref{Gamma_12}), 
this expression reduces to the form (\ref{P_exact}) given in the main text.

Lasing threshold is defined as the value of the temperature bias at which $N_d$, which satisfies Eq. (\ref{dot_Nd}), 
begins to exponentially grow in time provided it was small at the beginning. This  condition  can be formulated as
\begin{eqnarray}
\frac{d}{dN_d}\big[ \Gamma_{01}p_1 - \Gamma_{10}p_0 - \kappa_d (N_d - N_d^{\rm eq}) \big]\bigg|_{N_d\to 0}>0.
\nonumber
\end{eqnarray}
This condition can be simplified in the limit $\gamma_d\ll\gamma_h,\gamma_c$, which is  favorable for lasing.
In this limit we find
\begin{eqnarray}
\frac{N_h-N_c}{N_h(1+N_c)+(1+N_h)N_c+N_hN_c} > \frac{\kappa_d}{\gamma_d}.
\label{threshold}
\end{eqnarray}
In practical terms, since the combination in the left side of the equation does not exceed 1, one
should meet the conditions $\kappa_d\ll \gamma_d\ll \gamma_h,\gamma_c$ while designing the sample.

\section{Wigner function}

The Wigner function of the output resonator
is evaluated as
\begin{equation}
W(x',p')=\sum_{k,l} f_{kl}(x',p')\langle k|\rho^*_{\rm res}|l\rangle,
\label{W}
\end{equation}
where the sum runs over the states of the harmonic oscillator, 
the density matrix elements $\langle k|\rho^*_{\rm res}|l\rangle$ are found by solving
the Lindblad equation (\ref{mastereq}) and taking trace over all variables except the states of the output resonator,
and the function $f_{kl}(x',p')$ is defined as \cite{Wigner1932}
\begin{equation}
f_{kl}(x',p')=\int_{-\infty}^{\infty} \frac{dy}{2\pi} \psi_k^{*}\left(x'-\frac{y}{2} \right)\psi_l\left(x'+\frac{y}{2} \right) e^{-i y p'}.
\label{fkl}
\end{equation}
In Eqs. (\ref{W},\ref{fkl}) both the coordinate and the momentum are normalized by their zero point fluctuation values.
Next, $\psi_l(x)$ is the wave function of the harmonic oscillator corresponding to its energy level with the number $l$.
The integral (\ref{fkl}) can be solved exactly \cite{Groenewold1946}
\begin{eqnarray}
f_{kl}(x',p')&=& \frac{(-1)^k}{\pi}\sqrt{\frac{k!}{l!}}(4H)^{(l-k)}e^{-2H}e^{i(l-k)\arctan\frac{p'}{x'}}
\nonumber\\ && \times
L_l^{(l-k)}(4H),
\end{eqnarray}
where $H=(x'^2+p'^2)/2$ and $L_l^k(x')$ are the generalized Laguerre polynomials.

\end{document}